# Three-dimensional *operando* optical imaging of single particle and electrolyte heterogeneities inside Li-ion batteries

[1,2]Raj Pandya[✉], [1]Lorenzo Valzania[#], [3,4]Florian Dorchies[#], [1]Fei Xia[#], [5]Jeffrey Mc Hugh, [2,6]Angus Mathieson, [6]Jien Hwee Tan, [7]Thomas G. Parton, [6]Michael De Volder, [3,4]Jean-Marie Tarascon, [1]Sylvain Gigan[✉], [1]Hilton B. de Aguiar[✉], and [3,4]Alexis Grimaud[✉]

[1]Laboratoire Kastler Brossel, ENS-Université PSL, CNRS, Sorbonne Université, Collège de France, 24 Rue Lhomond, Paris 75005, France

[2]Department of Physics, Cavendish Laboratory, University of Cambridge, JJ Thomson Avenue, Cambridge CB3 0HE, United Kingdom

[3]Chimie du Solide et de l'Energie, UMR 8260, Collège de France, Paris, France

[4]Réseau sur le stockage Electrochimique de l'Energie (RS2E), Amiens, France

[5]Neuroglial Interactions in Cerebral Physiology and Pathologies, Center for Interdisciplinary Research in Biology, Collège de France, CNRS, INSERM, Labex Memolife, Université PSL, 75005 Paris, France

[6]Department of Engineering, University of Cambridge, Cambridge CB3 0FS, U.K.

[7]Yusuf Hamied Department of Chemistry, University of Cambridge, Lensfield Road, Cambridge, CB2 1EW, United Kingdom

[✉]Correspondence to: rp558@cam.ac.uk, sylvain.gigan@lkb.ens.fr, h.aguiar@lkb.ens.fr alexis.grimaud@college-de-france.fr

[#]These authors contributed equally

**Abstract**

**Understanding (de)lithiation heterogeneities in battery materials is key to ensuring optimal electrochemical performance and developing better energy storage devices. However, this remains challenging due to the complex three dimensional morphology of microscopic electrode particles, the involvement of both solid and liquid phase reactants, and range of relevant timescales (seconds to hours). Here, we overcome this problem and demonstrate the use of bench-top laser scanning confocal microscopy for simultaneous three-dimensional *operando* measurement of lithium ion dynamics in single particles, and the electrolyte, in batteries. We examine two technologically important cathode materials that are known to suffer from intercalation heterogeneities: $Li_xCoO_2$ and $Li_xNi_{0.8}Mn_{0.1}Co_{0.1}O_2$. The single-particle surface-to-core transport velocity of Li-phase fronts, and volume changes – as well as their inter-particle heterogeneity – are captured as a function of C-rate, and benchmarked to previous ensemble measurements. Additionally, we visualise heterogeneities in the bulk and at the surface of particles during cycling, and image the formation of spatially non-uniform concentration gradients within the liquid electrolyte. Importantly, the conditions under which optical imaging can be performed inside absorbing and multiply scattering materials such as battery intercalation compounds are outlined.**

**Introduction**

A key factor in enhancing the performance of Li-ion batteries is the development of high energy density cathode materials such as Ni-rich Lithium Nickel Manganese Cobalt Oxides (NMCs). A long debate still remains into the nature of ion (de)intercalation in such materials[1–4] with heterogeneities and irreversibilities in intercalation driving degradation and capacity fade[5–8]. Until such effects are thoroughly understood the necessary advances in battery performance are unlikely to be realised. However, one of the difficulties in probing ion (de)intercalation in battery electrodes, is the complex 3D morphology of constituent particles, with the particles microscopic surface, bulk and electrolyte environment all playing a role in (de)intercalation rates over seconds to hours[9–12].

In the life-sciences, optical reflection microscopies are a ubiquitous tool for low-cost, non-invasive, microscale characterisation of evolving systems[13,14]. Typically, microscopic reflection is performed at visible wavelengths (400 to 800 nm) either in the wide-field (WF) – where the entire sample is illuminated – or with confocal laser scanning (LSCRM), where a focussed laser beam is rapidly scanned across a given sample plane and the reflected light is used to build an image. A key difference between WF and LSCRM is that in the latter, specified depths of a material can be individually probed (sectioning). Over the last decade optical reflection microscopies, predominantly with WF or single plane illumination, have emerged as in-expensive tool for also tracking battery electrode dynamics[15–21].

This is because the reflectivity of many micron sized features on the electrode surface can dramatically increase, decrease or spectrally shift on (de)lithiation[22,23]. High numerical aperture (N.A.) microscope objectives have recently been used with the WF technique to extend its resolution and track changes in phase[24], state-of-charge[25] and structure[26] at the single-particle level in high energy density and fast charging battery materials. Despite this, little framework still exists for physically understanding the reflectivity signals from such (WF) experiments, with conflicting results emerging[24,27]. Furthermore, neither the limitations nor best-practices for optical reflection imaging of battery materials have been elucidated. Understanding how to apply the technique is particularly pertinent as the (real) refractive index (RI; $n$) of many battery electrode materials ($n$ = 2 to 3)[28] is strongly mismatched from that which high N.A. optical microscopies are designed for ($n$ ~1.4 to 1.6). The RI mismatch can result in severe aberrations and artifacts[29], which will be further accentuated by the high absorption and multiple scattering[30] of electrodes. In batteries specifically, high-resolution reflection microscopy studies have thus far been primarily limited to studying surface topography changes[19,20] or lithiation in 'flat' single crystalline particles in 2D[24–26]. Such constraints not only fundamentally limit applications but also skews our understanding as objects and processes are inherently three-dimensional (3D) with both a surface and bulk. Most importantly, as is the case with other *operando* methods such as X-Ray imaging, optical reflection methods have been unable to directly visualise the liquid electrolyte and its interaction with particles, making it challenging to completely resolve the origins of (de)intercalation heterogeneities.

Here, we overcome these limitations by first understanding, how, and under what conditions, optical imaging can be performed in batteries whilst avoiding false conclusions from optical artifacts. In contrast to other high-resolution optical microscopy studies, which have used 2D WF imaging, we focus on LSCRM, where sectioning allows single particles in 3D to be studied. We apply LSCRM to examine Li-intercalation in both Ni-rich NMC811 ($Li_xNi_{0.8}Mn_{0.1}Co_{0.1}O_2$) and LCO ($Li_xCoO_2$), both known to experience heterogeneities in Li occupancy during cycling, as characterised by other *operando* imaging techniques[10,31]. To benchmark the methods, single particle volume changes and surface-to-core phase front velocities during cycling are measured, with the inter-particle heterogeneity quantified and compared with results of ensemble studies. We further push the LSCRM technique to distinguish intercalation heterogeneities that occur in the bulk of particles from those limited to the surface, as well as visualise the formation of concentration gradients in the surrounding electrolyte, which we show can be simultaneously tracked from its intrinsic fluorescence. Our results highlight LSCRM as one of the only ways of microscopically imaging both the solid and liquid phases of batteries in 3D and in-*operando*.

**Main**

**Microscopic high-resolution optical imaging of metal oxide battery particles**

For many metal oxides, including LCO and NMC811, their reflectivity (and thus refractive index)[22,23,32] at visible wavelengths changes during insertion of a charge. To investigate the relationship between charge state and reflectivity, *ex-situ* microscopic reflection spectra of individual LCO and NMC811 particles are shown in Figure 1a as a function of lithiation ($x = 0$ corresponds to the fully delithiated particles). In both LCO and NMC811, the (surface) reflectivity increases with decreasing $x$ in the near-infrared between 680 and 900 nm for LCO and 800 and 900 nm for NMC811, with changes of up to 20% between $x \approx 0$ (0.5 for LCO) and $x \approx 1$. Hence, in this wavelength range the magnitude of the particle reflectivity can be used as a proxy of the state-of-charge. However, between 400-680 nm for LCO and 400-800 nm for NMC811, the changes in reflectivity with $x$ are non-monotonic, with spectral shifts consistent with bulk reflectivity measurements of other materials[28,33,34]. This region should consequently be avoided for microscopic measurements where reflectivity is used to track the lithiation state of particles in an electrode. Furthermore, in some common electrode materials, *e.g.* $Li_xFePO_4$, the reflectivity change (400 to 900 nm) at different lithiation stages is <1%, suggesting that microscopic reflectivity may not be a universal tool for tracking particle charge state (**supplementary information 1**) as recently proposed[24]. Fitting the reflection spectra in Figure 1a with the Kramers-Kronig relations[35] (see **supplementary information 2**) allows extraction of the state-of-charge dependent refractive index (RI). The real part of the refractive indices ($n$) for LCO and NMC811 is found to be between 2 – 2.5. This is larger than the RI (1.5 to 1.6) high numerical aperture (N.A.) microscope objective lenses are designed for, hence aberration corrections should be applied to account for the index mismatch (see discussion below).

To understand how *ex*-situ observations relate to changes in reflectivity during cycling, we perform *operando* reflection (confocal) microscopy measurements at several wavelengths on a single LCO particle (see methods and below), plotting the total reflected intensity and potential during charging in Figure 1b. At 800 nm (red curve), when light is focussed on/collected from the surface (S) of the particle, the reflectivity decreases linearly as $x$ is increased. Whereas between 380 nm and 600 nm a non-monotonic 'W' shaped curve is observed (green, blue, and purple curves) in line with Figure 1a. Interestingly, when focussing/collecting light from *inside* (I) the particle – an advantage offered by confocal sectioning – the reflectivity at 800 nm decreases linearly with $x$ (Figure 1b, blue dashed line). The difference between the 'S' and 'I' response at 800 nm can be reconciled by noting that the reflectivity at the interface of two media of RI $n_1$ and $n_2$ will be proportional to the difference, $(n_1-n_2)^2$ under normal incidence[36]. At the surface of particles, $n_1$ is equal to that of the electrolyte (~1.4 to 1.6)[37,38] and $n_2$ that of the particle 1.8-2.5[39]. However, inside the particle $n_1$ and $n_2$ will both vary between 1.8 and 2.5. Consequently, depending on the focus, different interfacial refractive indices, *i.e.*

electrolyte/particle *versus* particle/particle (intraparticle), will be resolved. We note that absorption of the particle top surface impacting layers below only results in attenuation of the 'I' response and cannot explain the opposing direction of trends between 'S' and 'I'. Indeed, the variation in the imaginary ($k$; absorptive) part of the refractive index is relatively small with lithiation state at 800 nm. Hence, the different extent to which $k$ (and its changes) contribute to the 'S' and 'I' signals during (de)lithiation are expected to be less significant than that of $n$ (see **supplementary information 2**). Our observations potentially explain the contrasting reports of Merryweather *et al.*[24] and Jiang *et al.*[27] with the former reporting brightening of LCO on delithiation and the latter dimming.

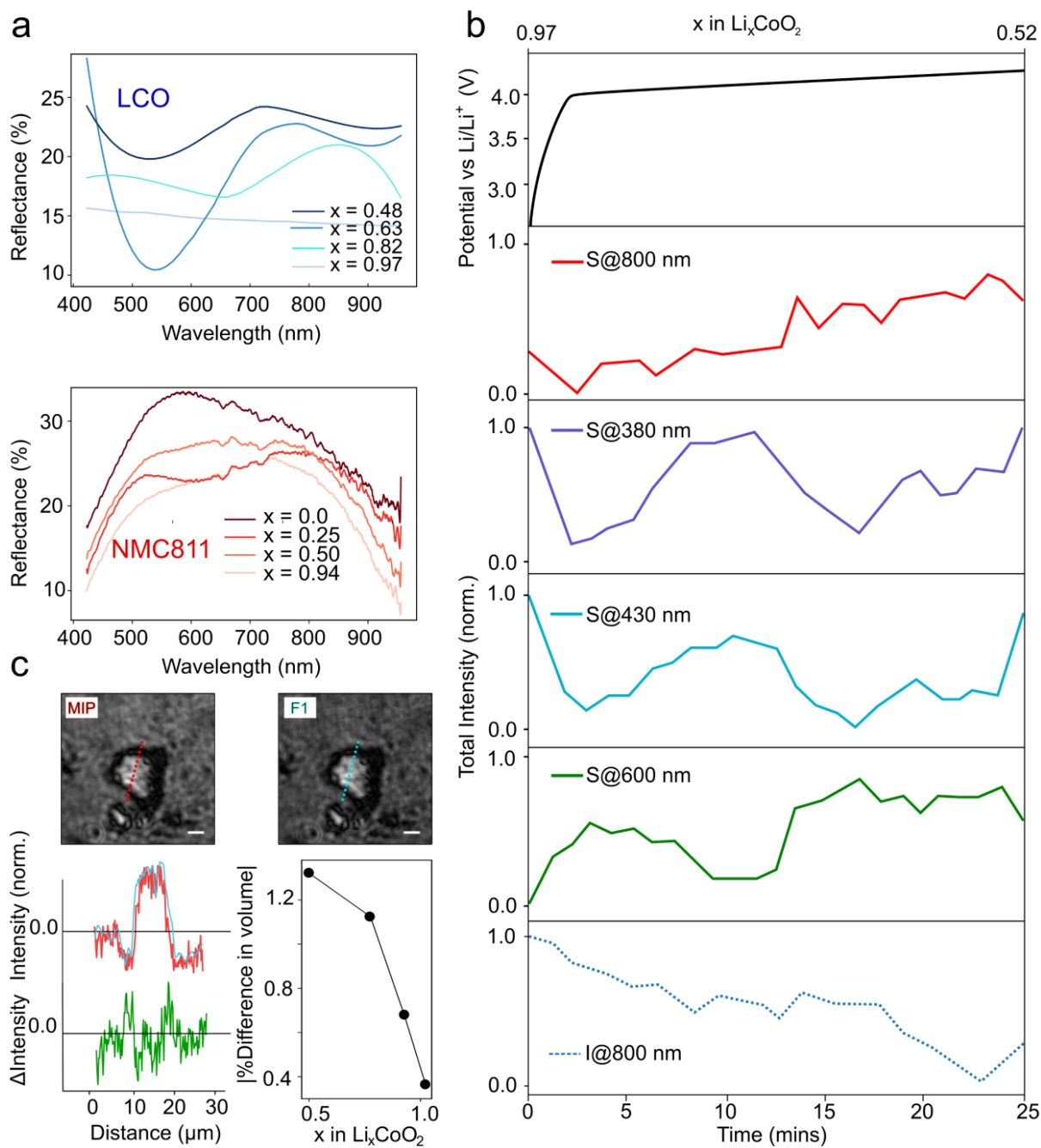

**Figure 1: Tracking charge state with optical reflection microscopy. a**. Reflectivity spectra of $Li_xCoO_2$ (LCO; blue; top) and $Li_xNi_{0.8}Mn_{0.1}Co_{0.1}O_2$ (NMC811; red; bottom) as a function of lithiation state ($x$). **b.** (Top) Spectral reflectivity as a function of time/state-of-charge ($x$) for single LCO particle during galvanostatic charge (dark blue curve below). S – focussing on particle surface (solid lines) and I – focussing inside particle (dashed line). Reflectivity measured from same particle at 800 nm (surface - red, inside blue), 600 nm (green), 430 nm (light blue), and 380 nm (purple). **c.** (Top) Maximum intensity projection (MIP) image of $Li_{0.05}CoO_2$ particle obtained from a defocus stack and image taken from a single focus position (F1). Scale bar 10 μm. (Left) Normalised line-cut of reflection intensity through particle (dashed red and blue lines in image) for MIP (red) and F1 (blue) images. Green line shows difference between two line-cuts. (Right) Absolute percentage difference in estimated particle volume between MIP and F1 as a function of state-of-charge ($x$) in $Li_xCoO_2$.

The results above suggest that the choice of the imaging plane plays a critical role when imaging such electrolyte/particle systems. Different combinations of the red and dotted blue curves in Figure 1b might be observed for different particles irrespective of any battery dynamics simply due to defocus or electrode roughness, *i.e.* inter-particle heterogeneities cannot be established from measurements at a single focal plane (WF). Additionally, interference between different reflection planes on the particle surface will influence the *spatial* distribution of intensity imaged. In **supplementary information 3 and supplementary video 1** we demonstrate (in pseudo-wide field measurements) that even in the absence of any cycling, different intensity patterns – closely resembling patterns that might be ascribed to the motion of phase-fronts or arise from morphology inhomogeneities[19,20] – can be observed across the surface of particles simply by small (20 nm) adjustments in the focus[40]. The large RI of the battery particles means only a small defocus, which may not be captured, or may even be created by, an auto-focus feedback system, can give rise to such effects. Hence, measuring at a range of focal planes, *i.e.* in 3D, is key to distinguish spatial intensity patterns modulated by small (high N.A amplified) focal shifts throughout a cycle, and true electrochemical dynamics.

The refractive index, $n$, also plays a role on the imaged size of an object *via* the optical path length, (OPL) OPL = $n \cdot s$, where $s$ is the geometric path length. Hence, before using optical microscopy to also investigate particle size changes during cycling[26], the impact of a varying $n$ on the OPL must first be accounted for[27]. In Figure 1c, we compare the lateral width of a LCO particle extracted from a reflection image at a single focus position (F1) and the lateral width of the same particle calculated from the maximum intensity projection[41–43] (MIP) where the influence of OPL on the object size is accounted for (see methods). By comparing the MIP and F1 images we can estimate an error in sizing introduced. For $x$ = 1 (in $Li_xCoO_2$), $L_{F1}$ and $L_{MIP}$ (L is the particle length; dashed red and blue lines in Figure 1c) deviate by 0.12%. This deviation increases monotonically with decreasing $x$, reaching 0.43% at $x$ = 0.48. From error propagations (see **supplementary information 4**) this will correspond to a deviation in volume of the imaged object (as compared to its 'true' size) of ~0.36% to 1.3% as $x$ decreases from 1 to 0.48 (right panel graph in Figure 1c and s**upplementary information 5**). The deviation will be more pronounced for smaller curved particles, with larger *n*, but generally our results suggest that changes in volume below ~1 to 1.5% are challenging to unambiguously detect with optical microscopy.

Having understood the limitations and conditions under which optical reflection imaging can be performed, we turn to applying these methods. LSCRM, as schematically depicted in Figure 2a, is performed on polycrystalline LCO (average particle diameter, $\langle d_{particle}\rangle \sim$ 2.5 to 7.5 μm) and NMC811 ($\langle d_{particle}\rangle$ ~5 μm). A laser beam at 820 nm, where absorption and scattering are minimised and changes in reflectivity with lithiation state are monotonic for LCO and NMC811, is scanned across the sample. Adjusting the focus of the beam in the sample, whilst using a pinhole (0.7 AU) to gate out-of-focus light, allows probing of different 2D planes up to the optical penetration length, which is <15 μm in both materials (see **supplementary information 6**). Stacking the 2D images and correcting for laser attenuation and defocus aberration (**supplementary information 6**) allows for sub-micrometre 3D information to be obtained on individual particles, as illustrated in Figure 2b,c. A customised battery half-cell with optical access (right side of Figure 2a) allows for galvanostatic cycling during the recording of confocal image stacks of electrodes (~120 s per stack). All data presented herein, is taken from the first 2 to 9 cycles of an electrode.

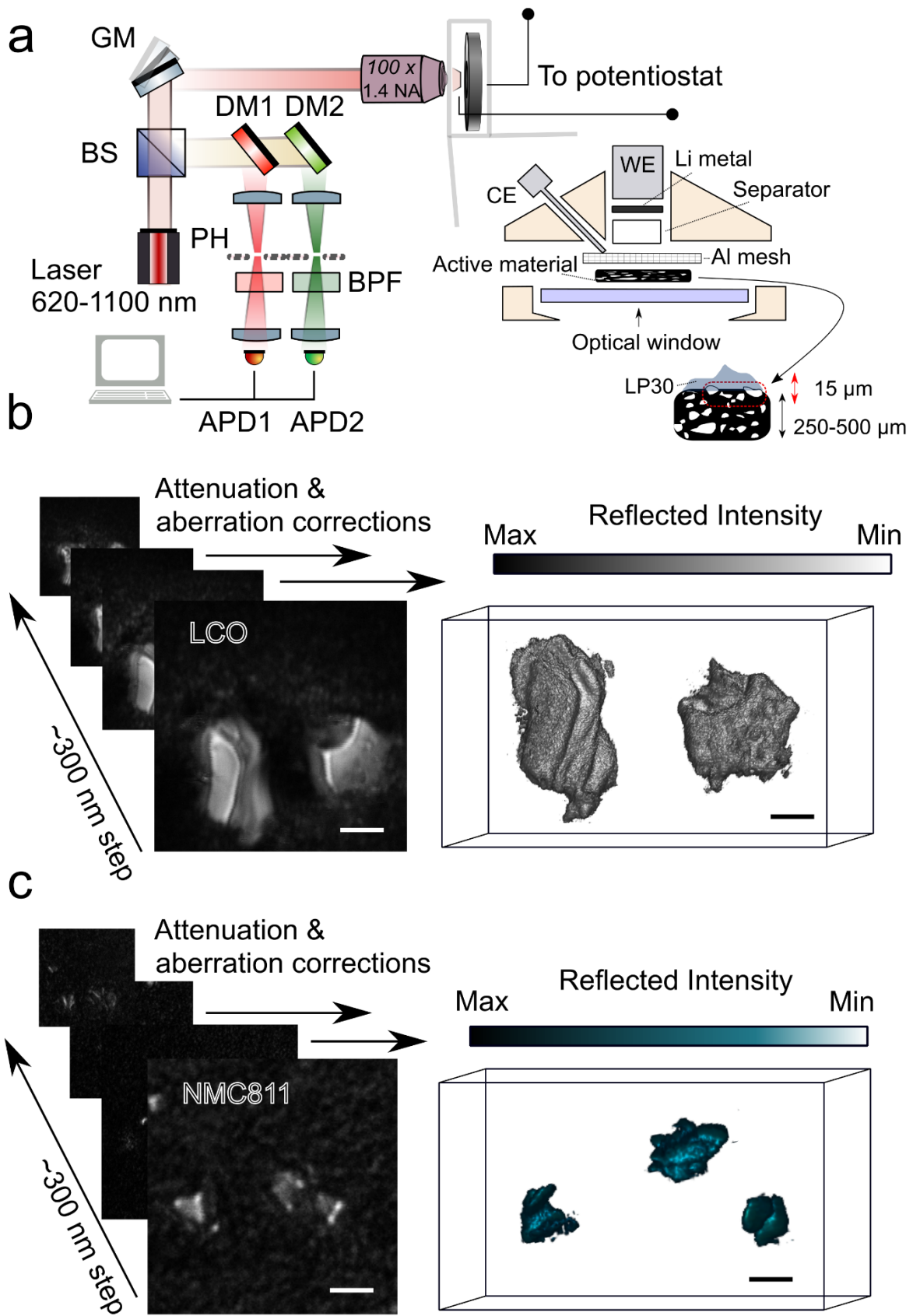

**Figure 2: Single particle three-dimensional imaging in polycrystalline battery electrodes. a.** Cartoon schematic of laser scanning confocal microscopy (LSCRM) setup and *operando* battery cell. A tunable laser source is directed to microscope body *via a* beam-splitter (BS) with reflection signal

and fluorescence passed through pinholes (PH) before being collected onto two separate avalanche photodiodes (APDs). Long pass dichroic mirrors (DM) and band pass filters (BPF), control the spectral selectivity; APD1: 700-1100 nm and APD2: 550-680 nm. z-sampling is performed by movement of an objective piezo. Self-standing LCO and NMC811 electrodes are placed in an optical microscopy half-cell (WE, working electrode; CE, counter electrode). A lithium metal counter is used along with a glass fibre separator with wetting using carbonate liquid electrolyte (LP30 (see methods); blue shading); depth resolution (red arrow) is limited to ~15 μm. **b-c.** Confocal z-stacks of LCO and NMC811. Scale bar 5 μm. Following attenuation correction[44,45], aberration correction[29] and thresholding (see **supplementary information 6**) a 3D reconstruction of the particle can be obtained. z-sampling rate is pre-determined as ~300 nm.

*Operando* tracking of single particle volume changes

To first benchmark our LSCRM methodology, we monitor volume changes of individual particles during a cycle. By calculating the number of voxels occupied by particle containing regions within the reconstructed volumes, volumes at the single particle level can be obtained. In Figure 3a-c, the normalised change in single particle volumes ($\Delta V/V$) of LCO and NMC811 are shown as a function of time/charge state and C-rate. The volume of LCO particles in all cases increases during charging and decreases during discharging (standard deviation in maximum volume change ($\sigma_{\Delta V/V}$) is 0.42), with a maximum volume change of ~3% ±1% (see **supplementary information 6** for error bar estimation). For NMC811, $\sigma_{\Delta V/V}$ is 0.63 with the volume decreasing during the charge ($|\Delta V/V_{max}|$ ~12% ±2%), as shown by the graph in Figure 3c. In NMC811 we observe significant changes in volume only between lithium fractions of $x = 0.3$ and $x = 0.7$ and as for LCO there are also inactive particles which show a $\Delta V/V$ ~0 throughout the cycle. In NMC811 there is a greater scattering of results with a hysteresis in $\Delta V/V$, suggesting irreversible changes, *e.g.* primary particle cracking or movement within the secondary particle microstructure. However, for polycrystalline secondary particles (NMC811) and polycrystalline particles (LCO) as studied here, contributions from insertion of lithium into the lattice, cracking and rearrangement cannot be disentangled[46]. Indeed, for a small fraction (~15%) of NMC811 particles measured, positive changes in $\Delta V/V$ are observed (see **supplementary information 6)**. Nevertheless, the results in Figure 3b,c are consistent both in terms of magnitude, direction and state-of-charge onset, for volume changes obtained by previous ensemble X-ray diffraction and pressure dependent open circuit voltage measurements of LCO ($\Delta V/V_{x=1 \rightarrow x=0.5}$ ~ +2%) and NMC811 ($\Delta V/V_{x=1 \rightarrow x=0.2}$ ~ -8%) [46–49].

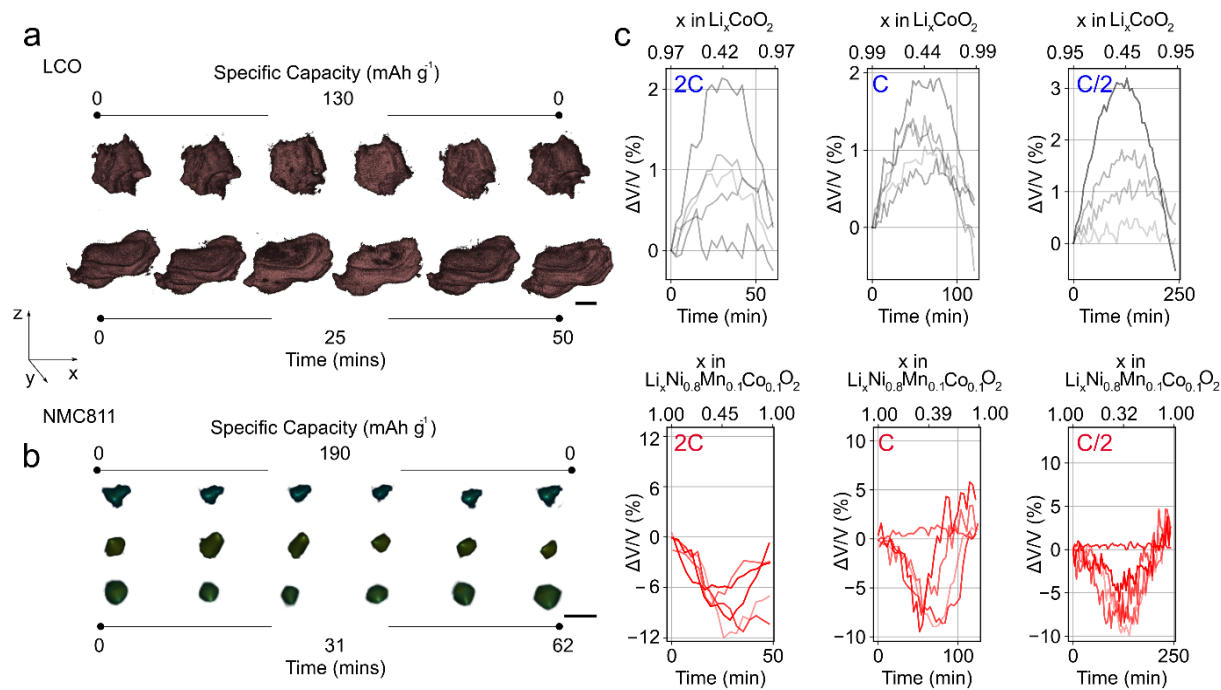

**Figure 3: Tracking particle volume changes during cycling a-b.** Tomographic reconstruction of LCO and NMC811 throughout a 2C charge with specific capacities and times above and below. Scale bar is 5 μm. **c.** Percentage change in particle volume as compared to that at the open circuit potential (OCP; black LCO and red NMC811) as a function of time during a 2C, C and 0.5C charge-discharge sequence; $x$ in $Li_xCoO_2$ and $Li_xNi_{0.8}Mn_{0.1}Co_{0.1}O_2$ plotted above. Different grey and red curves represent volumes changes of different particles to highlight the large degree of heterogeneity in volume changes.

**Surface-to-core transport**

To further understand how structural inhomogeneities might be related to (de)lithiation asymmetries, we examine the transport velocities of Li-ion containing phase fronts from the exterior to the centre of particles. The surface-to-core transport velocity is indeed a parameter of fundamental interest as local inter- and intra-particle heterogeneities in ion transport can limit the overall (dis)charging rate and potentially cause irreversible material changes[50–52]. To track the motion of phase fronts through polycrystalline NMC811 or LCO particles during the cycle, the normalised time varying reflectivity, a proxy for the state-of-charge, is extracted for each z-plane across the central area of a particle. From this, the depth-dependence of the time-varying reflectivity can be extracted, as shown in Figure 4a, b as a function of cycling rate. For each z-plane, the point at which the reflectivity crosses zero is then determined (dashed line in Figure 4a, b); from the variation of this reference point with depth, a transport velocity for ion containing fronts can be estimated (see **supplementary information 7**).

For NMC811, in Figure 4a, the delithiation and lithiation both occur from the surface to the core, in a *quasi* symmetric manner across the particle z-span, with qualitatively similar behaviours from C/2 to 2C. This observation is in-line with a shrinking core type mechanism of ion (de)intercalation[1,53,54], where, because of the higher lithium flux on the particle surface as compared to the bulk, ion (de)insertion is diffusion limited, rather than surface limited[55]. In LCO, delithiation occurs from the surface to the core, as for NMC811. However, lithiation occurs *quasi* uniformly across the particle volume, with all depths changing reflectivity simultaneously (flat dashed line in Figure 4b). These observations match phase-field modelling[56] and several experimental studies[1,57,58] which have previously suggested that, in contrast to delithiation, lithiation of LCO is charge transfer limited and occurs *via a* lithium-poor phase with higher ionic diffusion, resulting in a intercalation wave type mechanism. This differing mechanism of ion transport may explain the uniform surface-to-core lithiation profile and significantly higher diffusivity on lithiation than delithiation for our LCO.

Because the velocity of phase fronts ($v_p$) will depend on the state-of-charge, values extracted in Figure 4c-e represent an average across lithiation/delithiation, albeit at the single particle level. In NMC811 and LCO, $v_p$ increases with C-rate, but remains of the same order of magnitude of 2-6 nm s$^{-1}$ across C-rates for delithiation and lithiation. Note that for LCO, velocities cannot be extracted during lithiation as it is beyond our time resolution (100 s). These values sit at the lower end of those reported previously in the literature (1 nm s$^{-1}$ to 50 nm s$^{-1}$)[24,59–64]. However, in this work, the transport velocities reported are *through* individual polycrystalline particles *i.e.* from the surface to the core, which cannot be obtained with other methods such as galvanostatic intermittent titration technique (GITT) or 2D imaging which operate at the ensemble level and/or do not have 3D directional resolution on transport. We note that anisotropy in the transport is not expected due to the random orientation of primary particles and their polycrystalline nature (see **supplementary information 7**). Finally, the linear scaling of $v_p$ with C-rate in both NMC811 and LCO suggests that phase transport is kinetically rather than thermodynamically limited for both layered oxides[10,31].

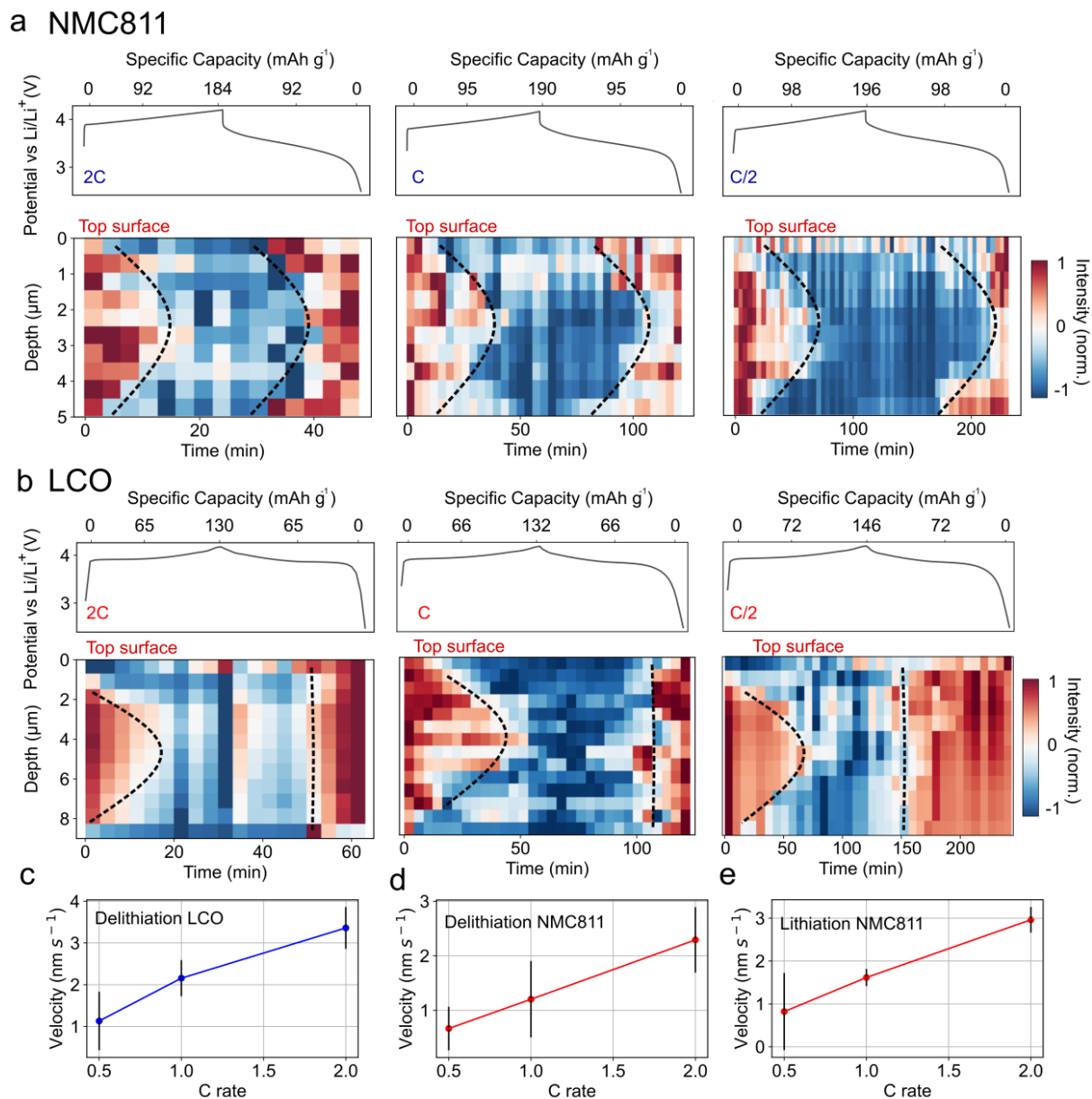

**Figure 4: Measurement of phase front velocities through single particles. a-b.** Reflectivity as a function of z-plane and charge state/time through a LCO and NMC811 particle, respectively. The top of the particle (0 μm) is taken as the first plane inside the particle. Galvanostatic charge-discharge is performed at C rates from 2C to 0.5C (top panels); all data is taken after first cycle. Dashed line is guide to the eye at which normalised ([-1,1]) reflected intensity changes sign. From the time/depth dependence of this point a phase front velocity through the particle can be estimated (see **supplementary information 7**). **c-e.** Phase front velocity for delithiation and lithiation (only NMC811) as a function of C-rate; spread obtained from measurement on 9 NMC811 and 7 LCO particles.

**Imaging phase-front dynamics on different surfaces**

Having shown that the velocity of phase fronts from the surface to the core of particles can be tracked using LSCRM, we push further the technique to spatially separate (de)lithiation inhomogeneities between the surface and bulk of single particles. For that, we compare using mathematical reconstructions, at different points during a charge-discharge cycle, the spatial distribution of reflected intensity (lithiation state) between the particle exterior, and core for LCO particles cycled at different C-rates (C/2 and 2C). For both the exterior and core the reflection contrast is derived from refractive index changes within the material (see **supplementary information 8**). LCO particles are first computationally 'unwrapped' into shells (~25) of thickness ~500 nm for each time point in the cycle. The minimum volume ellipsoid – a close approximation of the particle shape – enclosing points of the shells is then calculated[65,66]. The surface of the ellipsoid is projected onto a 2D plane using the Mercator projection[67], as show in Figure 5a (see **supplementary information 8**), and colours correspond to different phases or domains with different lithiation states. In Figure 5b,c, 2D projections of the outer (exterior) and inner (core) most shells are shown for pristine, charged and discharged states during galvanostatic cycles at C/2 and 2C. At C/2, for the particle exterior, the contrast in reflectivity both within and between domains remains small throughout the cycle indicating large area, uniform (de)lithiation. Furthermore, for the pristine and discharged states the distribution of different lithiation domains is near-identical, demonstrating a reversible cycling process. At the particle centre, a spatial rearrangement of lithium phases is observed from the pristine to charged states. This rearrangement remains on discharging, but the differences in lithiation degree between domains, for a given overall state-of-charge, are small. The observation of changes at both the particle exterior and core however indicate that (de)lithiation occurs throughout the entire particle.

For 2C, at the particle exterior, the pristine and discharged states do not show a similar spatial distribution of lithiation phases. For the particle core, no rearrangement of lithium domains at the end of charge is observed, unlike at C/2. Only minimal changes in both the contrast and spatial distribution of lithium domains occur between the pristine, charged and discharged states. This observation suggests that (de)lithiation does not occur throughout the entire particle volume, in agreement with the limited cycling capacity measured at 2C when compared to C/2 (see Figure 3b for instance). For the exterior, regions of contrastingly high (dark blue) and low (yellow) Li-content appear on going from the pristine to charged state, and persist to the discharged state. This indicates non-uniform lithiation on the surface of LCO particles at high C-rate, consistent with several previous observation in layered oxide materials[11,68–71].

Our results do suggest that LSCRM, combined with mathematical/computational treatments, can spatially distinguish phase inhomogeneities simultaneously at different surfaces within a single cycling particle. However, for such a non-trivial technique heavily relying on data treatment, more remains to understand the implication of these observation, with for instance NMC811 showing a much more complex behaviour (see **supplementary information 9**).

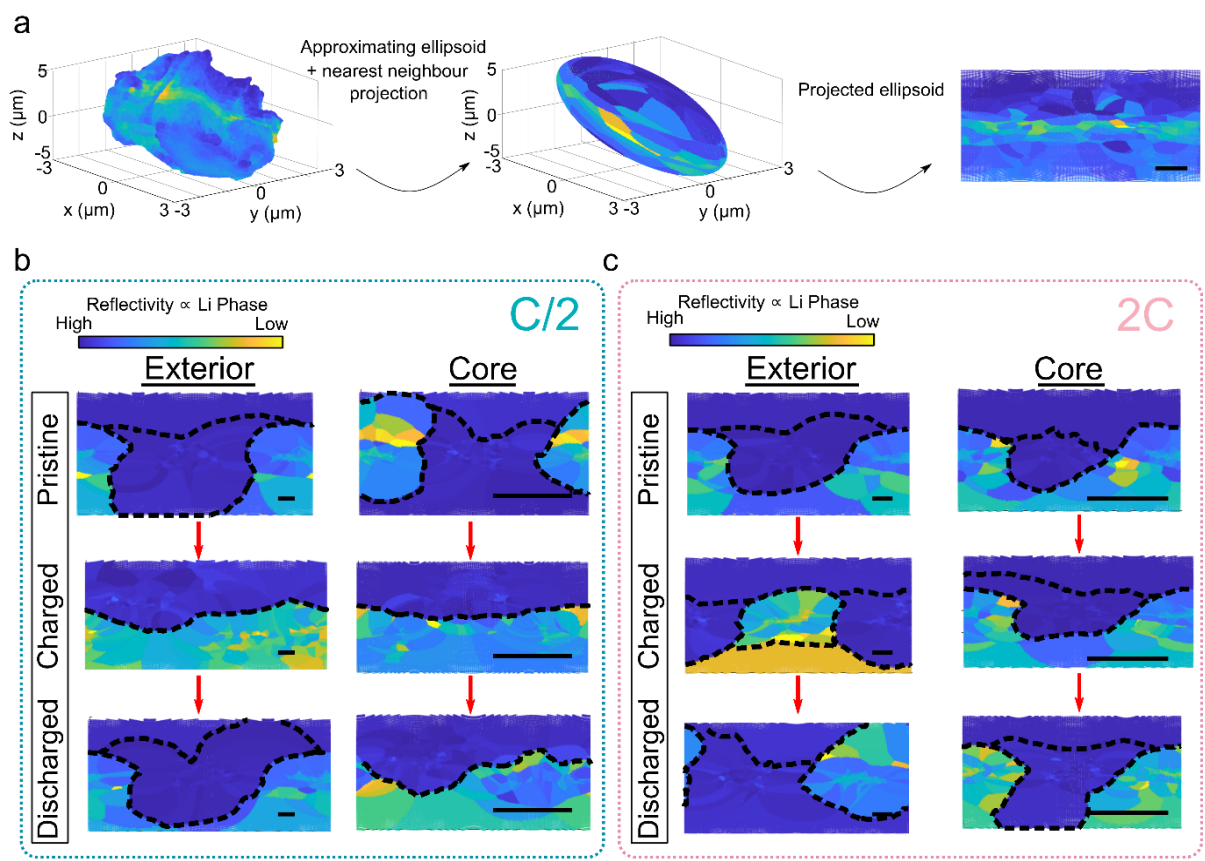

**Figure 5: Unwrapping of particle surfaces a.** Cartoon schematic demonstrating enclosing of particle 'shells' onto the surface of an ellipsoid followed by projection onto the 2D plane. **b-c.** 2D projections from shells at exterior and centre of LCO particles in pristine ($x = 0.96$ at C/2 and $x = 0.95$ at 2C in in $Li_xCoO_2$), charged ($x = 0.44$ at C/2 and $x = 0.52$ at 2C) and discharged ($x = 0.95$ at C/2 and $x = 0.94$ at 2C) states for C/2 (b) and 2C (c) charge-discharge cycle. Dashed lines act as guide to show region of different Li-phases (colour scale). Scale bar in each image is 5 μm. The ellipsoidal shell that encloses the exterior surface has semi-major axis radii of 6.5, 6.5 and 7.3 μm whereas for the core it is 1, 1 and 1.8 μm.

**Visualisation of electrolyte dynamics**

Thus far, optical and other *operando* imaging methods such X-Ray microscopy have been limited to visualising particles of active material. However, we can show that LSCRM offers the opportunity to image both lithium transport in solids, as demonstrated above, simultaneously with concentration gradients forming in the liquid electrolyte upon polarisation. Indeed, $LiPF_6$, the Li-conducting salt in the electrolyte used, has previously been observed to be a source of fluorescence at visible wavelengths, with the exact origin albeit debated (see discussion in Laurence *et al.*[72]). Following these past observations, we find that under two-photon excitation (2PEF), the fluorescence of $LiPF_6$ salt (solid) is significantly more efficient than with one photon (1PEF), particularly above an excitation wavelength of 900 nm (Figure 6a). In Figure 6b we perform ex-*situ* 2PEF experiments on $LiPF_6$ solutions (in 1:1 vol.% ethylene carbonate/dimethyl carbonate) at different concentrations (ranging from 0.01 to 4.0 mol/L), and find a strong monotonous brightening with concentration. Figure 6c shows the evolution of the electrolyte 2PEF intensity, at different focal planes above a self-standing LCO electrode, during a charge-discharge cycle at 2C. In all cases the electrolyte 2PEF increases near linearly on charging to 4.2 V, *i.e.* on the release of Li-ions to the electrolyte, with a slight plateauing in the rate of brightening between 4.1 V – 4.2 V. Discharging, *i.e.* depletion of lithium from the electrolyte, results in dimming of the electrolyte 2PEF. The response is repeatable and of a similar magnitude over 8 cycles (see **supporting information 10**). Together the ex-*situ* and *operando* measurements show the electrolyte 2PEF to be highly sensitive to the Li-ion concentration and indicate that the 2PEF must indeed derive from the $LiPF_6$ salt or a compound bound to it[72]. Furthermore, the measurements indicated that the 2PEF is unrelated to cycling induced electrolyte degradation[73–77]. Consequently, it is suggested that the intensity of the 2PEF can be used to qualitatively track the Li-ion concentration in the electrolyte.

Our attention is then turned to examining the spatial distribution of salt concentration upon cycling. To do so, LSCRM experiments are repeated with an excitation of 1020 nm with two separate detectors, one for collecting particle reflectance and another for 2PEF (see Figure 2a; ~250 s per confocal stack). This allows for simultaneous volumetric imaging of the electrolyte and LCO particles, as illustrated in Figure 6d. In Figure 6e, the 2PEF signal of the electrolyte in the plane ~300 nm above the top surface of the electrodes is plotted at selected time points (labelled A to F; correspond to points of equivalent state-of-charge as marked in Figure 6c) and spatial locations during a C/2 cycle (dashed black lines indicate regions of similar 2PEF intensity). Upon charge, the 2PEF intensity increases due to delithiation and an increase in the $LiPF_6$ concentration. A gradient in the 2PEF intensity also becomes present above potentials of 4.0 V around LCO particles. This 2PEF/electrolyte concentration gradient, which decays *quasi* homogeneously away from particles, originates from the difference between the rate of delithiation at the LCO surface and the rate at which $PF_6^-$ anions diffuse towards the LCO particles[78–81] to balance $Li^+$ ions released and maintain electroneutraility[74,82,83]. Furthermore, at all potentials the 2PEF is relatively uniform in regions not containing LCO (bottom row) and upon discharge, the concentration gradient around particles disappears. Altogether, our results indicate a reversible process, as would indeed be expected for the formation of a concentration gradient around active material upon cycling. The concentration gradient we observe extends as far as 1.5 μm. This in agreement with previous theoretical and experimental studies which have shown electrolyte concentration gradients around electrode interfaces extending between 500 nm[84] and 10s μm[74,79,85] depending on the electrode, electrolyte composition (mass to volume ratio) and cycling conditions[86,87]. We note that on charging the electrode and allowing it to relax to OCP, the polarisation gradient rapidly disappears (see **supporting information 10**). Switching to a greater C-rate, *i.e.* 2C, during the initial charge up to 3.9 V (A panels), the 2PEF of the electrolyte is brightest around LCO particles with the intensity decaying *quasi* homogeneously away from the particles. Increasing the potential to 4.2 V (B, C, D and E) results in a drastic increase in 2PEF intensity but the distribution of electrolyte 2PEF also becomes heterogeneous around both particles and in regions of the electrode ~20 – 50 μm away from LCO particles (bottom row of Figure 6f). The latter observation suggests inhomogeneous electrolyte diffusion within the self-standing electrode, as a result of the geometry and distribution of pores within the particle-carbon/binder matrix[88]. This influence of the matrix can be expected at higher C-rates where the rate of (de)lithiation at the LCO surface and ionic diffusion rate in the electrolyte are more significantly mismatched[89]. Finally, on discharging to 3.9 V (panel F) the 2PEF distribution returns to its initial state, confirming that our observations do not stem from electrolyte degradation but are from (de)lithiation and local changes in salt concentration and concentration gradients within the electrolyte. In NMC811, similar behaviour is observed as for LCO but with subtle differences requiring further in-depth analysis (see **supporting information 10**).

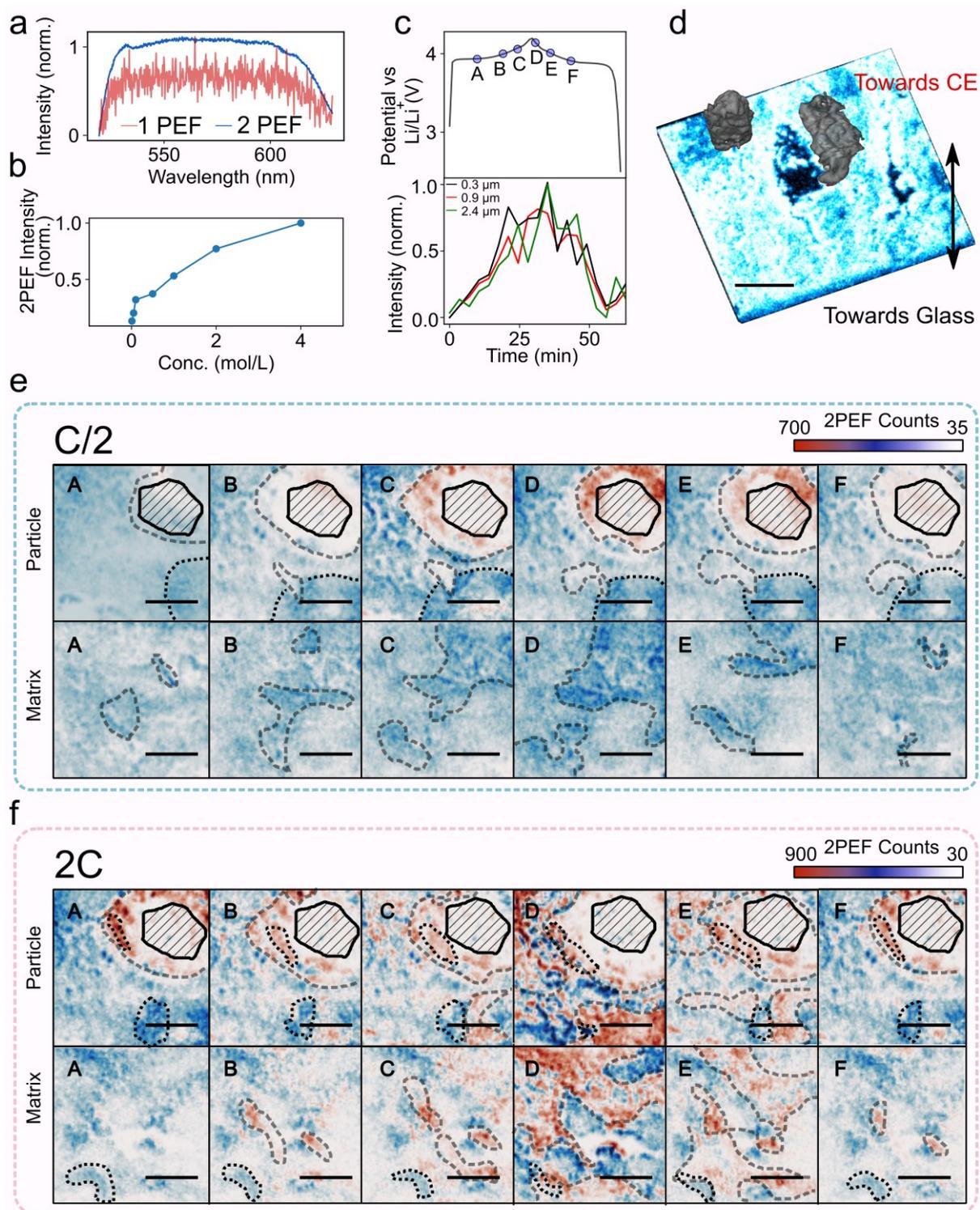

**Figure 6: Two photon excited fluorescence from battery electrolytes. a.** Two-photon excited (2PEF; 1020 nm excitation; black) and one-photon excited (1PEF; 532 nm; red) fluorescence spectra of LiPF$_6$ solid; a laser filter at 532 nm prevents resolution of the spectrum below this wavelength. We find the 2PEF to be significantly brighter than the 1PEF in LiPF$_6$, but this is challenging to quantify. **b.** Normalised maximum 2PEF intensity as function of LiPF$_6$ concentration in 1:1 vol.% ethylene carbonate/dimethyl carbonate. **c.** Charge-discharge curve of electrode at 2C plotted as a function of charge state (top). Corresponding change in electrolyte 2PEF as a function of time/charge-state (bottom) at planes 0.3, 0.9 and 2.4 μm above the 'surface' of particles. **d.** 3D volumetric reconstruction of electrolyte (blue, fluorescence) volume sitting above an electrode with particles embedded in the electrolyte (grey, reflectance). We only image electrolyte surrounding/above the top plane particles *i.e.* towards the glass observation window. Scale bar is 5 μm. **e-f.** 2PEF from plane directly above particles (marked in black with fill lines) for electrode cycled at C/2 (d) and 2C (e). 2 regions of interest are shown: surrounding a particle and a region containing only the host matrix. Dashed grey lines highlight regions of local 2PEF intensity enhancement. Dotted black lines are guide lines for regions whose spatial position of 2PEF does not evolve significantly throughout the cycle. Scale bar is 4 μm.

**Conclusion and Outlook**

This paper demonstrates that high-resolution LSCRM is a powerful tool for 3D microscopic tracking of single particle structural transformations, Li-ion intercalation and electrolyte dynamics in operating batteries. Applying LSCRM to LCO and NMC811 important functional parameters, such as the velocity of Li-containing phase fronts from the surface to the core of individual particles and the single-particle volume change during a cycle, are obtained. These parameters are shown to agree well with ensemble measurements, but also display a large degree of interparticle heterogeneity not captured by ensemble techniques. A unique advantage of LSCRM is the ability in-*operando* to simultaneously visualise electrolyte concentration gradients and lithiation heterogeneities at the particle surface and particle bulk in 3D, setting it aside from not only other optical methods, but also X-Ray and electron microscopies.

Further work measuring even more particles of different size[90] and crystallinity[91] is needed to place our material specific observations, such as the rapid lithiation observed in our LCO, on firmer footing. Although $LiPF_6$ is the source of intrinsic electrolyte fluorescence, particular attention is needed to unravel the exact compound responsible, with both trace organic molecules bound to $LiPF_6$ and halophosphates derivatives suggested to be responsible[72,77]. The influence of multiple coordinated Li-species on the 2PEF intensity must also be taken into account such that a more quantitative assessment can be performed[92], as well as methods to separate salt and solvent transport. Correlating the electrolyte distribution more precisely with the activity of particle surfaces is crucial to further disentangle the contribution of both to (de)intercalation asymmetries. Understanding this interplay between particles and electrolyte is even more critical for reaction mechanisms encompassing conversion[93] or alloying[94] for which particle cracking and/or volume expansion lead to electrolyte decomposition upon cycling. Limitations remain in-terms of depth, speed and chemical specificity of the optical methods. However, using mid-infrared photo-thermal techniques[95,96], wavefront shaping, and/or fibre optics to achieve imaging at greater depths (and in full-cells)[97,98]; wide-field schemes such as full field optical-coherence tomography[99] or spinning-disk confocal[14] to improve the speed; and more chemically sensitive probes such as the Raman effect[100,101] will further push the utility of the technique and complement existing synchrotron imaging methods. Nonetheless, the label-free nature of LSCRM, outlined application framework and limited need to develop specialised equipment[13] means beyond batteries such methods will find use in the study of 3D solid and liquid phase dynamics in a range of functional materials and devices ranging from electrocatalysts[102] to bioelectronics[103].

## Data Availability

All data is made freely available at [url to be added in proof].

## Author Contributions

R.P conceived the idea, performed optical experiments, analysed and interpreted the data and wrote the manuscript. L.V developed the ellipsoid projection code and along with F.X. developed, analysed and interpreted the optical tomography experiments. F.D. prepared *operando* cells and interpreted the data. J.M supervised all confocal microscopy measurements. T.G.P performed *ex*-situ reflection microscopy experiments. A.M., J.H.T and M.D.V provided experimental advice and interpreted the data. J.M.T. interpreted the data. S.G. and H.B.A supervised the project and interpreted the optical data. A.G. supervised the project, interpreted the data and wrote the manuscript. All authors contributed to preparation of the manuscript.

## Acknowledgements

R.P acknowledges financial support from Clare College, University of Cambridge and thanks (all Cambridge): Arjun Ashoka for insightful discussions on optical microscopy, Scott Keene for critical reading of the manuscript, Christoph Schnedermann and Alice J. Merryweather for advice on preparation of *operando* cells and Prof. Ulrich F. Keyser for loan of a potentiostat and spectrometer. L.V. acknowledges funding from the Swiss National Science Foundation (grant P400P2_199329). F.D acknowledges the École normale supérieure Paris-Saclay for his PhD scholarship. T.G.P. acknowledges the ESPRC NanoDTC (EP/L015978/1). The authors thank Prof. Nathalie Rouach (Collège de France) and Prof. Silvia Vignolini (Cambridge) for access and use of experimental resources.

## Supplementary Information

## Methods

## Reagents

Carbon black Super P Conductive (99+% metals basis, Alfa Aesar), ethylene carbonate/dimethyl carbonate 1:1 vol.% (Dodochem), $LiCoO_2$ (99.8% trace metals basis, Sigma Aldrich), $LiFePO_4$ carbon coated (≥99.5%, battery grade, Sigma-Aldrich), $LiPF_6$ (under argon, 99.9%, Solvionic), 1M $LiPF_6$ in 1:1 vol.% ethylene carbonate/dimethyl carbonate (LP30, Dodochem), 60 wt % polytetrafluoroethylene dispersion in water (Sigma-Aldrich) and $LiPF_6$ solid (99.9%, battery grade, Sigma-Aldrich) were purchased and used without additional treatment. The synthesis of polycrystalline NMC811 particles was adapted from literature[104] to yield ~ 5 μm particles with Ni, Mn and Co content ratios of 0.8, 0.1 and 0.1 respectively.

## Preparation of LCO self-standing electrodes

50.0, 112.5 and 145.8 mg of polycrystalline LCO, carbon black Super P conductive and 60 wt% polytetrafluoroethylene (PTFE) dispersion in water were carefully mixed and grinded in a mortar with a pestle for 15 minutes. Ethanol was added from time to time in order to help the different components bind together and structure the mixture. When the mixture became thick it was taken out of the mortar and rolled on a clean surface for 15min in order to make a smooth and ~ 0.5 mm thick electrode film with LCO, carbon and PTFE mass ratios of 20, 45 and 35% respectively. The as-prepared film was dried overnight in an 80°C oven under vacuum.

**Preparation of NMC811 self-standing electrodes**

50.0, 112.5 and 87.5 mg of polycrystalline NMC811, carbon black Super P conductive and dry PTFE were carefully mixed and grinded in a mortar with a pestle for 15 minutes in an argon-filled glovebox. Few drops of acetonitrile were added from time to time in order to help the different components bind together and structure the mixture. When the mixture became thick it was taken out of the mortar and rolled on a clean surface inside the glovebox for 15min in order to make a smooth and ~ 0.5 mm thick electrode film with NMC811, carbon and PTFE mass ratios of 20, 45 and 35% respectively.

**Preparation of materials for *ex*-situ studies**

Lithium Cobalt Oxide, Lithium Nickel Manganese Cobalt Oxide and Lithium Iron Phosphate (LFP)

LCO, NMC811 and LFP powder synthetized or purchased as previously described above were ball-milled for 20 minutes with 10% in weight of carbon Super P. Swagelock cells were assembled in Ar-filled glovebox with approximatively 10 mg of powder, using two whattman separators filled with LP30 electrolyte and with metallic lithium as counter electrode. Cells were cycled with fixed capacity, before powder was recovered and washed two times with DMC, and then centrifuged and dried for 1 hour under vacuum. This allowed for the preparation of LCO, NMC811 and LFP powders with different states of charge/lithium fraction for *ex*-situ experiments.

Lithium hexafluorophosphate solutions in 1:1 vol.% ethylene carbonate/dimethyl carbonate

In an argon-filled glovebox, 3.1, 15.2, 30.4, 151.9, 303.8, 607.6 and 1215.2 mg of $LiPF_6$ were dissolved in 2.0 mL of 1:1 vol.% ethylene carbonate/dimethyl carbonate (EC/DMC) mixture to respectively yield 0.01, 0.05, 0.1, 0.5, 1.0, 2.0 and 4.0 mol.$L^{-1}$ $LiPF_6$ solutions in EC/DMC.

***Ex*-situ reflection microspectroscopy**

Reflection microspectroscopy of individual battery particles was performed on a customised Zeiss Axio microscope. Illumination was provided using a halogen lamp (Zeiss HAL100) focussed by a 50x/0.4 objective (Nikon, T Plan SLWD). Reflected light was spatially filtered (collection spot diameter <5 μm) using a 100 μm-diameter optical fibre (Avantes FC-UV100-2-SR) mounted in confocal configuration and connected to a spectrometer (Avantes AvaSpec-HS2048). Particles were pressed onto glass microscope slides before measurement and encapsulated with a second cover glass inside any Ar glovebox to avoid aerial oxidation/hydrolysis. A minimum of 20 individual particles (well separated from any carbon) were used to obtain spectra.

**Particle size error estimation using time-domain optical coherence tomography**

Time-domain optical coherence tomography (TD-OCT) was used to generate the MIP and F1 images discussed in the main text. F1 was taken from the position in the time-gate position where there was maximum electric field. For the TD-OCT setup, a beam delivered from a Ti:sapphire laser (MaiTai HP, Spectra-Physics) was divided into two paths by a polarizing beam splitter. On one path the light was focused onto the back focal plane of a 0.8 N.A. objective (Olympus) which delivers collimated light on the sample. The reflected light from the sample is collected by a beam splitter and recombined with the second path on another beam splitter. Light on this second (reference) arm does not pass onto the sample but onto a delay stage (Newport) which controls temporal overlap between reference and signal arms. The combined signal and reference light are imaged onto is a charged coupled device camera (Manta G-046B, Allied Vision). A polarizer in front of the camera ensures that only light with a selected polarization is measured. The reference arm is scanned to measure the electric field amplitude as a function of time-delay (focal positon) within the sample. From this trace and the MIP and F1 images can be generated.

***Ex*-situ one and two-photon excited fluorescence spectroscopy**

For liquid measurements 1 mm path length glass cuvettes were filled with Lithium hexafluorophosphate solutions in 1:1 vol.% ethylene carbonate/dimethyl carbonate of a predefined concentration. Cuvettes were sealed in an Argon containing glove box before measurement. For measuring the solid $LiPF_6$, powder was sandwiched between two microscope coverslides. The coverslides were sealed together with epoxy resin inside an Argon glove box before measurement. For one photon excitation light from a 532 nm laser (Laser Quantum Gem, CW, 532 nm, 100 mW) was focussed on to the sample with a 50 mm focal length lens (Thorlabs). The fluorescence was collected collimated and focussed onto the fibre port of a spectrometer (Ocean Optics, Ventanna 532 nm). The exciting laser light was separated from the fluorescence using a dichroic mirror (Semrock 532 nm RazorEdge). For two photon excited fluorescence experiments the same configuration was used except the exciting light was the 1020 nm output of a Ti:sapphire laser (MaiTai HP, Spectra-Physics) and an additional 1000 nm short pass filter (Thorlabs) was placed in front of the spectrometer to remove any of the exciting beam. We note that the 2PEF efficiency dramatically decreased with wavelength and when exciting below 900 nm (450 nm one photon excited fluorescence) very little emission was observed. For 1PEF the laser power at the sample was ~200 mW and for 2PEF the power at the sample was ~145 mW.

*Operando* **Laser Scanning Confocal Microscopy**

Laser scanning confocal microscopy was performed using a custom-built microscope. The output of a Chameleon Ultra II Ti:Sapphire laser (Coherent), was directed to a laser (galvo-) scanning microscope body (Scientifica), with x-y-z piezo control. The reflected light from the sample was collected and focussed through a 0.7 A.U pinhole for spatial filtering. Dichroic mirrors (500-620 nm and >650 nm) spectrally filtered the light which was focussed onto two silicon avalanche photodiodes (APDs). Recording was performed using the Abberior Instruments Imspector 16 software. Depending on the exact experiment and signal magnitude the pixel dwell time was varied between 10 and 20 μs, with regions of between 300 × 300 and 600 × 600 pixles scanned again depending on the particles of interest. Pixels ranged between 40 and 80 nm in size. Overall this resulted in z-stack acquisition times of between 100 and 220s. The coherence length of the laser used here is <2 μm such that reflection interference contrast effects can be minimised. In all experiments a laser power at the sample of <200 mW was used. For experiments shown in shown in Figure 1b LEDs at the appropriate excitation wavelength were used as the source.

The *operando* half-cell (ECC-Opto-Std; El-Cell GmBH) was modified to accommodate a 1.4 N.A objective and 0.15 – 0.17 μm thickness coverslips (see **supporting information 11**). All cycling was performed using a Gamry Reference 600 potentiostat, with homebuilt software to control synchronisation between the potentiostat and microscope. Throughout the manuscript a C-rate of 1 C corresponds to a charge in 1 h.

**Image processing**

Image processing was performed with custom Python, Matlab and ImageJ[105] scripts (see **supporting information 2 to 7** for further details of algorithms). Before performing any analysis or image correction, xyz image registration was performed using the ImageJ registration plugin[106].

**References**

1.  Fraggedakis, D. *et al.* A scaling law to determine phase morphologies during ion intercalation. *Energy Environ. Sci* **13**, 2142 (2020).

2.  Van Der Ven, A., Ceder, G., Asta, M. & Tepesch, P. D. First-principles theory of ionic diffusion with nondilute carriers. *Phys. Rev. B* **64**, (2001).

3.  Van Der Ven, A., Bhattacharya, J. & Belak, A. A. Understanding Li Diffusion in Li-Intercalation Compounds. *Acc. Chem. Res.* **46**, 1216–1225 (2013).

4.  Panchmatia, P. M., Armstrong, A. R., Bruce, P. G. & Islam, M. S. Lithium-ion diffusion mechanisms in the battery anode material Li1+xV1-xO2. *Phys. Chem. Chem. Phys.* **16**, 21114–21118 (2014).

5.  Pender, J. P. *et al.* Electrode Degradation in Lithium-Ion Batteries. *ACS Nano* **14**, 1243–1295 (2020).

6.  Shang, M., Chen, X. & Niu, J. Nickel-rich layered LiNi 0.8 Mn 0.1 Co 0.1 O 2 with dual gradients on both primary and secondary particles in lithium-ion batteries. *Cell Reports Phys. Sci.* **3**, (2022).

7.  Liu, T. *et al.* Correlation between manganese dissolution and dynamic phase stability in spinel-based lithium-ion battery. *Nat. Commun.* **10**, (2019).

8.  Liu, T. *et al.* Rational design of mechanically robust Ni-rich cathode materials via concentration gradient strategy. *Nat. Commun.* **12**, 1–10 (2021).

9.  Daemi, S. R. *et al.* Visualizing the Carbon Binder Phase of Battery Electrodes in Three Dimensions. *ACS Appl. Energy Mater.* **1**, 3702–3710 (2018).

10. Park, J. *et al.* Fictitious phase separation in Li layered oxides driven by electro-autocatalysis. *Nat. Mater.* **20**, 991–999 (2021).

11. Mistry, A., Heenan, T., Smith, K., Shearing, P. & Mukherjee, P. P. Asphericity Can Cause Nonuniform Lithium Intercalation in Battery Active Particles. *ACS Energy Lett.* **7**, 1871–1879 (2022).

12. Finegan, D. P. *et al.* Spatial quantification of dynamic inter and intra particle crystallographic heterogeneities within lithium ion electrodes. *Nat. Commun.* **11**, 1–11 (2020).

13. Jonkman, J., Brown, C. M., Wright, G. D., Anderson, K. I. & North, A. J. Tutorial: guidance for quantitative confocal microscopy. *Nat. Protoc. 2020 155* **15**, 1585–1611 (2020).

14. Qin, S., Isbaner, S., Gregor, I. & Enderlein, J. Doubling the resolution of a confocal spinning-disk microscope using image scanning microscopy. *Nat. Protoc. 2020 161* **16**, 164–181 (2020).

15. Arai, H., Yaguchi, A., Nishimura, Y., Akimoto, Y. & Ikezawa, A. Operando Optical Analysis of LiFePO4 Composite Electrodes. *J. Phys. Chem. C* **125**, 3776–3780 (2021).

16. White, E. R., Lodico, J. J. & Regan, B. C. Intercalation events visualized in single

microcrystals of graphite. *Nat. Commun.* **8**, 1–9 (2017).

17. Beaulieu, L. Y., Cumyn, V. K., Eberman, K. W., Krause, L. J. & Dahn, J. R. A system for performing simultaneous in situ atomic force microscopy/optical microscopy measurements on electrode materials for lithium-ion batteries. *Rev. Sci. Instrum.* **72**, 3313 (2001).

18. Otoyama, M., Sakuda, A., Hayashi, A. & Tatsumisago, M. Optical microscopic observation of graphite composite negative electrodes in all-solid-state lithium batteries. *Solid State Ionics* **323**, 123–129 (2018).

19. Chen, Y. *et al.* Operando video microscopy of Li plating and re-intercalation on graphite anodes during fast charging. *J. Mater. Chem. A* **9**, 23522–23536 (2021).

20. Sanchez, A. J., Kazyak, E., Chen, Y., Lasso, J. & Dasgupta, N. P. Lithium stripping: anisotropic evolution and faceting of pits revealed by operando 3-D microscopy. *J. Mater. Chem. A* **9**, 21013–21023 (2021).

21. Liu, F. *et al.* Dynamic spatial progression of isolated lithium during battery operations. *Nature* **600**, 659–663 (2021).

22. Wu, W., Wang, M., Ma, J., Cao, Y. & Deng, Y. Electrochromic Metal Oxides: Recent Progress and Prospect. *Adv. Electron. Mater.* **4**, (2018).

23. Gillaspie, D. T., Tenent, R. C. & Dillon, A. C. Metal-oxide films for electrochromic applications: present technology and future directions. *J. Mater. Chem.* **20**, 9585–9592 (2010).

24. Merryweather, A. J., Schnedermann, C., Jacquet, Q., Grey, C. P. & Rao, A. Operando optical tracking of single-particle ion dynamics in batteries. *Nature* **594**, 522–528 (2021).

25. Xu, C. *et al. Operando visualisation of kinetically-induced lithium heterogeneities in single-particle layered Ni-rich cathodes*. doi:10.26434/chemrxiv-2022-qb80n

26. Merryweather, A. J. *et al. Operando monitoring of single-particle kinetic state-of-charge heterogeneities and cracking in high-rate Li-ion anodes*. (2021). doi:arXiv:2111.11997

27. Jiang, D. *et al.* Optical Imaging of Phase Transition and Li-Ion Diffusion Kinetics of Single LiCoO2 Nanoparticles During Electrochemical Cycling. *J. Am. Chem. Soc.* **139**, 186–199 (2017).

28. Joshi, Y. *et al.* Modulation of the Optical Properties of Lithium Manganese Oxide via Li-Ion De/Intercalation. *Adv. Opt. Mater.* **6**, (2018).

29. Diel, E. E., Lichtman, J. W. & Richardson, D. S. Tutorial: avoiding and correcting sample-

induced spherical aberration artifacts in 3D fluorescence microscopy. *Nat. Protoc. 2020 159* **15**, 2773–2784 (2020).

30. Rohrbach, A. Artifacts resulting from imaging in scattering media: a theoretical prediction. *Opt. Lett.* **34**, 3041–3043 (2009).

31. Grenier, A. *et al.* Intrinsic Kinetic Limitations in Substituted Lithium-Layered Transition-Metal Oxide Electrodes. *J. Am. Chem. Soc.* **142**, 7001–7011 (2020).

32. Ong, G. K. *et al.* Electrochromic Niobium Oxide Nanorods. *Chem. Mater.* **32**, 468–475 (2020).

33. Liu, H. L. *et al.* Electronic structure and lattice dynamics of Li x CoO 2 single crystals. *New J. Phys.* **17**, (2015).

34. Beluze, L. *et al.* Infrared electroactive materials and devices. *J. Phys. Chem. Solids* **67**, 1330–1333 (2006).

35. Kuzmenko, A. B. Kramers-Kronig constrained variational analysis of optical spectra. *Rev. Sci. Instrum.* **76**, 083108 (2005).

36. Mahmoodabadi, R. G. *et al.* Point spread function in interferometric scattering microscopy (iSCAT). Part I: aberrations in defocusing and axial localization. *Opt. Express* **28**, 25969–25988 (2020).

37. Jin, Y. *et al.* In operando plasmonic monitoring of electrochemical evolution of lithium metal. *Proc. Natl. Acad. Sci. U. S. A.* **115**, 11168–11173 (2018).

38. Kitta, M., Murai, K., Yoshii, K. & Sano, H. Electrochemical Surface Plasmon Resonance Spectroscopy for Investigation of the Initial Process of Lithium Metal Deposition. *J. Am. Chem. Soc.* **143**, 11160–11170 (2021).

39. Muñoz-Castro, M. *et al.* Controlling the optical properties of sputtered-deposited LixV2O5 films. *J. Appl. Phys* **120**, 135106 (2016).

40. Contreras-Naranjo, J. C., Silas, J. A. & Ugaz, V. M. Reflection interference contrast microscopy of arbitrary convex surfaces. *Appl. Opt.* **49**, 3701–3712 (2010).

41. Volume Rendering in Three-Dimensional Display of SPECT Images. *J. Nucl. Med.* **8**, 1421–1428 (1990).

42. Hormel, T. T. *et al.* Maximum value projection produces better en face OCT angiograms than mean value projection. *Biomed Opt Express.* **9**, 6412-6424. (2018).

43. Badon, A. *et al.* Smart optical coherence tomography for ultra-deep imaging through highly

scattering media. *Sci. Adv.* **2**, (2016).

44. Gong, P. *et al.* Parametric imaging of attenuation by optical coherence tomography: review of models, methods, and clinical translation. *J. Biomed. Opt* **25**, 40901 (2020).

45. Ghosh, B., Mandal, M., Mitra, P. & Chatterjee, J. Attenuation corrected-optical coherence tomography for quantitative assessment of skin wound healing and scar morphology. *J. Biomed. Opt* **25**, 40901 (2020).

46. Koerver, R. *et al.* Chemo-mechanical expansion of lithium electrode materials – on the route to mechanically optimized all-solid-state batteries. *Energy Environ. Sci.* **11**, 2142–2158 (2018).

47. Wang, Y., Liu, J., Chen, T., Lin, W. & Zheng, J. Factors that affect volume change during electrochemical cycling in cathode materials for lithium ion batteries. *Phys. Chem. Chem. Phys.* **24**, 2167–2175 (2022).

48. Kondrakov, A. O. *et al.* Anisotropic Lattice Strain and Mechanical Degradation of High- and Low-Nickel NCM Cathode Materials for Li-Ion Batteries. *J. Phys. Chem. C* **121**, 3286–3294 (2017).

49. Lim, J. M. *et al.* Intrinsic Origins of Crack Generation in Ni-rich $LiNi_{0.8}Co_{0.1}Mn_{0.1}O_2$ Layered Oxide Cathode Material. *Sci. Reports 2017 71* **7**, 1–10 (2017).

50. Leanza, D. *et al.* Revealing the Dual Surface Reactions on a HE-NCM Li-Ion Battery Cathode and Their Impact on the Surface Chemistry of the Counter Electrode. *ACS Appl. Mater. Interfaces* **11**, 6054–6065 (2019).

51. Li, S. *et al.* Mutual modulation between surface chemistry and bulk microstructure within secondary particles of nickel-rich layered oxides. *Nat. Commun.* **11**, 1–9 (2020).

52. Tian, C., Lin, F. & Doeff, M. M. Electrochemical Characteristics of Layered Transition Metal Oxide Cathode Materials for Lithium Ion Batteries: Surface, Bulk Behavior, and Thermal Properties. *Acc. Chem. Res.* **51**, 89–96 (2018).

53. Palmer, M. G. *et al.* In situ phase behaviour of a high capacity $LiCoPO_4$ electrode during constant or pulsed charge of a lithium cell. *Chem. Commun.* **52**, 14169–14172 (2016).

54. Pritzker, M. D. Shrinking-core model for systems with facile heterogeneous and homogeneous reactions. *Chem. Eng. Sci.* **51**, 3631–3645 (1996).

55. Singh, G. K., Ceder, G. & Bazant, M. Z. Intercalation dynamics in rechargeable battery materials: General theory and phase-transformation waves in $LiFePO_4$. *Electrochim. Acta* **53**, 7599–7613 (2008).

56. Bazant, M. Z. Theory of chemical kinetics and charge transfer based on nonequilibrium thermodynamics. *Acc. Chem. Res.* **46**, 1144–1160 (2013).

57. Jow, T. R., Delp, S. A., Allen, J. L., Jones, J.-P. & Smart, M. C. Factors Limiting Li + Charge Transfer Kinetics in Li-Ion Batteries . *J. Electrochem. Soc.* **165**, A361–A367 (2018).

58. Dahéron, L. *et al.* Electron transfer mechanisms upon lithium deintercalation from LiCoO 2 to CoO2 investigated by XPS. *Chem. Mater.* **20**, 583–590 (2008).

59. Geng, Z., Chien, Y. C., Lacey, M. J., Thiringer, T. & Brandell, D. Validity of solid-state Li+ diffusion coefficient estimation by electrochemical approaches for lithium-ion batteries. *Electrochim. Acta* **404**, 139727 (2022).

60. Wei, Y. *et al.* Kinetics Tuning of Li-Ion Diffusion in Layered Li(NixMnyCoz)O2. *J. Am. Chem. Soc.* **137**, 8364–8367 (2015).

61. Märker, K., Reeves, P. J., Xu, C., Griffith, K. J. & Grey, C. P. Evolution of Structure and Lithium Dynamics in LiNi0.8Mn0.1Co0.1O2 (NMC811) Cathodes during Electrochemical Cycling. *Chem. Mater.* **31**, 2545–2554 (2019).

62. Xie, J. *et al.* Orientation dependence of Li-ion diffusion kinetics in LiCoO2 thin films prepared by RF magnetron sputtering. *Solid State Ionics* **179**, 362–370 (2008).

63. Xia, H., Lu, L. & Ceder, G. Li diffusion in LiCoO2 thin films prepared by pulsed laser deposition. *J. Power Sources* **159**, 1422–1427 (2006).

64. Ashton, T. E. *et al.* Stoichiometrically driven disorder and local diffusion in NMC cathodes. *J. Mater. Chem. A* **9**, 10477–10486 (2021).

65. Todd, M. J. & Yildirim, E. A. On Khachiyan's algorithm for the computation of minimum-volume enclosing ellipsoids. *Discret. Appl. Math.* **155**, 1731–1744 (2007).

66. Khachiyan, L. G. Rounding of Polytopes in the Real Number Model of Computation. *Math. Oper. Res.* **21**, 307–320 (1996).

67. Maling, D. H. *Coordinate systems and map projections*. (Pergamon Press, 1992).

68. Cogswell, D. A. & Bazant, M. Z. Theory of coherent nucleation in phase-separating nanoparticles. *Nano Lett.* **13**, 3036–3041 (2013).

69. Gent, W. E. *et al.* Persistent State-of-Charge Heterogeneity in Relaxed, Partially Charged Li1−xNi1/3Co1/3Mn1/3O2 Secondary Particles. *Adv. Mater.* **28**, 6631–6638 (2016).

70. Mu, L. *et al.* Propagation topography of redox phase transformations in heterogeneous layered oxide cathode materials. *Nat. Commun. 2018 9 1* **9**, 1–8 (2018).

71. Tan, C. *et al.* Nanoscale state-of-charge heterogeneities within polycrystalline nickel-rich layered oxide cathode materials. *Cell Reports Phys. Sci.* **2**, 100647 (2021).

72. Laurence, S. & Hardwick, J. Kerr gated Raman spectroscopy of LiPF 6 salt and LiPF 6-based organic carbonate electrolyte for Li-ion batteries. *Phys. Chem. Chem. Phys.* **21**, 23833

73. Kamiya, J., Mitsui, T., Ooe, M. & Sato, K. New Process of Synthesizing LiPF6 in Organic Solvent for Electrolyte. *ECS Meet. Abstr.* **MA2015-02**, 136 (2015).

74. Cheng, Q. *et al.* Operando and three-dimensional visualization of anion depletion and lithium growth by stimulated Raman scattering microscopy. *Nat. Commun. 2018 91* **9**, 1–10 (2018).

75. Jarry, A. *et al.* The Formation Mechanism of Fluorescent Metal Complexes at the LixNi0.5Mn1.5O4−δ/Carbonate Ester Electrolyte Interface. *J. Am. Chem. Soc* **137**, 31 (2015).

76. Yu, Y. *et al.* Coupled LiPF6 Decomposition and Carbonate Dehydrogenation Enhanced by Highly Covalent Metal Oxides in High-Energy Li-Ion Batteries. *J. Phys. Chem. C* **122**, 27368–27382 (2018).

77. Aurbach, D., Moshkovich, M., Cohen, Y. & Schechter, A. The Study of Surface Film Formation on Noble-Metal Electrodes in Alkyl Carbonates/Li Salt Solutions, Using Simultaneous in Situ AFM, EQCM, FTIR, and EIS. *Langmuir* **15**, 2947–2960 (1999).

78. Wang, A. A. *et al.* Potentiometric MRI of a Superconcentrated Lithium Electrolyte: Testing the Irreversible Thermodynamics Approach. *ACS Energy Lett.* **6**, 3086–3095 (2021).

79. Fawdon, J., Ihli, J., Mantia, F. La & Pasta, M. Characterising lithium-ion electrolytes via operando Raman microspectroscopy. *Nat. Commun. 2021 121* **12**, 1–9 (2021).

80. Fawdon, J., Rees, G. J., Mantia, F. La & Pasta, M. Insights into the Transport and Thermodynamic Properties of a Bis(fluorosulfonyl)imide-Based Ionic Liquid Electrolyte for Battery Applications. *J. Phys. Chem. Lett* **13**, 40 (2022).

81. Klett, M. *et al.* Quantifying mass transport during polarization in a Li Ion battery electrolyte by in situ 7Li NMR imaging. *J. Am. Chem. Soc.* **134**, 14654–14657 (2012).

82. Seo, D. M., Borodin, O., Han, S.-D., Boyle, P. D. & Henderson, W. A. Electrolyte Solvation and Ionic Association II. Acetonitrile-Lithium Salt Mixtures: Highly Dissociated Salts. *J. Electrochem. Soc.* **159**, A1489–A1500 (2012).

83. Brissot, C., Rosso, M., Chazalviel, J. -N. & Lascaud, S. In Situ Concentration Cartography in the Neighborhood of Dendrites Growing in Lithium/Polymer-Electrolyte/Lithium Cells. *J. Electrochem. Soc.* **146**, 4393–4400 (1999).

84. Holtz, M. E. *et al.* Nanoscale imaging of lithium ion distribution during in situ operation of battery electrode and electrolyte. *Nano Lett.* **14**, 1453–1459 (2014).

85. Richardson, G. W., Foster, J. M., Ranom, R., Please, C. P. & Ramos, A. M. Charge transport modelling of Lithium-ion batteries. *Eur. J. Appl. Math.* 1–49 (2021).

86. Uchida, S. & Kiyobayashi, T. How does the solvent composition influence the transport properties of electrolyte solutions? LiPF6 and LiFSA in EC and DMC binary solvent. *Phys. Chem. Chem. Phys.* **23**, 10875–10887 (2021).

87. Wang, A. A., Hou, T., Karanjavala, M. & Monroe, C. W. Shifting-reference concentration cells to refine composition-dependent transport characterization of binary lithium-ion electrolytes. *Electrochim. Acta* **358**, 136688 (2020).

88. Khan, Z. A., Agnaou, M., Sadeghi, M. A., Elkamel, A. & Gostick, J. T. Pore Network Modelling of Galvanostatic Discharge Behaviour of Lithium-Ion Battery Cathodes. *J. Electrochem. Soc.* **168**, (2021).

89. Kang, J., Koo, B., Kang, S. & Lee, H. Physicochemical nature of polarization components limiting the fast operation of Li-ion batteries. *Chem. Phys. Rev.* **2**, 041307 (2021).

90. Luo, Y. *et al.* Effect of crystallite geometries on electrochemical performance of porous intercalation electrodes by multiscale operando investigation. *Nat. Mater. 2021 212* **21**, 217–227 (2021).

91. Li, J. *et al.* Comparison of Single Crystal and Polycrystalline LiNi 0.5 Mn 0.3 Co 0.2 O 2 Positive Electrode Materials for High Voltage Li-Ion Cells. *J. Electrochem. Soc.* **,164**, A1534–A1544 (2017).

92. Bruce, P. G., Hardgrave, M. T. & Vincent, C. A. The determination of transference numbers in solid polymer electrolytes using the Hittorf method. *Solid State Ionics* **53**–**56**, 1087–1094 (1992).

93. Wang, F. *et al.* Conversion reaction mechanisms in lithium ion batteries: Study of the binary metal fluoride electrodes. *J. Am. Chem. Soc.* **133**, 18828–18836 (2011).

94. Zhou, L. *et al.* Electrolyte Engineering Enables High Stability and Capacity Alloying Anodes for Sodium and Potassium Ion Batteries. *ACS Energy Lett.* **5**, 766–776 (2020).

95. Zhao, J. *et al. Bond-Selective Intensity Diffraction Tomography*. (2022). doi:arXiv:2203.13630

96. Bai, Y., Yin, J. & Cheng, J. X. Bond-selective imaging by optically sensing the mid-infrared photothermal effect. *Sci. Adv.* **7**, 1559–1573 (2021).

97. Blanquer, L. A. *et al.* Optical sensors for operando stress monitoring in lithium-based batteries containing solid-state or liquid electrolytes. *Nat. Commun.* **13**, (2022).

98. Ganguli, A. *et al.* Embedded fiber-optic sensing for accurate internal monitoring of cell state in advanced battery management systems part 2: Internal cell signals and utility for state estimation. *J. Power Sources* **341**, 474–482 (2017).

99. Assayag, O. *et al.* Large Field, High Resolution Full-Field Optical Coherence Tomography. *Technol Cancer Res Treat.* **13**, 4558–4546 (2014).

100. Cui, Q. H. *et al.* Asymmetric Photon Transport in Organic Semiconductor Nanowires through Electrically Controlled Exciton Diffusion. *Sci. Adv.* **4**, (2018).

101. Flores, E., Novák, P. & Berg, E. J. In situ and Operando Raman spectroscopy of layered transition metal oxides for Li-ion battery cathodes. *Front. Energy Res.* **6**, 82 (2018).

102. Pavlovic, Z., Ranjan, C., Gao, Q., Van Gastel, M. & Schlögl, R. Probing the structure of a water-Oxidizing anodic iridium oxide catalyst using raman spectroscopy. *ACS Catal.* **6**, 8098–8105 (2016).

103. Rivnay, J. *et al.* Structural control of mixed ionic and electronic transport in conducting polymers. *Nat. Commun. 2016 71* **7**, 1–9 (2016).

104. Pimenta, V. *et al.* Synthesis of Li-Rich NMC: A Comprehensive Study. *Chem. Mater.* **29**, 9923–9936 (2017).

105. Schneider, C. A., Rasband, W. S. & Eliceiri, K. W. NIH Image to ImageJ: 25 years of image analysis. *Nat. Methods 2012 97* **9**, 671–675 (2012).

106. Lowe, D. G. Distinctive Image Features from Scale-Invariant Keypoints. *Int. J. Comput. Vis.* **60**, 91–110 (2004).